\documentstyle[12pt]{article}
\begin{document}
\title{Effects of Varying $G$}
\author{B.G. Sidharth$^*$\\
Centre for Applicable Mathematics \& Computer Sciences\\
B.M. Birla Science Centre, Adarsh Nagar, Hyderabad - 500 063 (India)}
\date{}
\maketitle
\footnotetext{$^*$Email:birlasc@hd1.vsnl.net.in; birlard@ap.nic.in}
\begin{abstract}
We consider two cosmologies which give an inverse time dependance for the
gravitational constant $G$, and show that we can recover the correct value
of the perhelion precession and anomalous inward radial accelerations,
amongst other things.
\end{abstract}
\section{Introduction}
In certain cosmological schemes, as is well known, the universal constant of
gravitation $G$, changes very slowly with time\cite{r1,r2,r3,r4}. These
include Dirac's large number cosmology and fluctuational cosmology. In these
cases the variation is given by
\begin{equation}
G = \frac{\beta}{T}\label{e1}
\end{equation}
where in fluctuational cosmology referred to (cf. also refs.\cite{r5,r6}), $\beta$
is given in terms o f the constant microphysical parameters by
$$\beta \approx \frac{l \hbar}{m^2}$$
where $l$ is the pion Compton wavelength and $m$ its mass.\\
(It may be pointed out in passing that while Dirac's cosmology has well known
inconsistencies, the latter theory is consistent with observation and predicts
an ever expanding universe as latest observations of distant supernovae do
indeed confirm. In addition, not only are the Large Number coincidences accounted
for, but also Weinberg's mysterious empirical relation between the pion mass and
the Hubble constant is deduced from the theory (cf. references).)\\
In any case, what we now propose to show is, that starting from (\ref{e1}),
we can account for the perhelion precession of the planet Mercury as also for
anomalous acceleration of the planets and more generally for the solar system
bodies and in addition predict anomalous changes in the orbital eccentricities.\\
\section{Solar System Orbits}
We now deduce using (\ref{e1}), the perhelion precession of Mercury. We first
observe that from (\ref{e1}) it follows that
\begin{equation}
G = G_o (1+ \frac{t}{t_o})\label{e2}
\end{equation}
where $G_o$ is the present value of $G$ and $t_o$ is the present age
of the universe and $t$ the time elapsed from the present epoch. Similarly one
could deduce that (cf.ref.\cite{r1}),
\begin{equation}
r = r_o \left(\frac{t_o}{t_o+t}\right)\label{e3}
\end{equation}
We next use Kepler's Third law\cite{r7}:
\begin{equation}
\tau = \frac{2 \pi a^{3/2}}{\sqrt{GM}}\label{e4}
\end{equation}
$\tau$ is the period of revolution, $a$ is the orbit's semi major axis,
and $M$ is the mass
of the sun. Denoting the average angular velocity of the planet by
$$\dot \Theta \equiv \frac{2 \pi}{\tau},$$
it follows from (\ref{e2}), (\ref{e3}) and (\ref{e4}) that
$$\dot \Theta - \dot \Theta_o = - \dot \Theta_0 \frac{t}{t_o},$$
where the subscript $o$ refers to the present epoch, \\
Whence,
\begin{equation}
\omega (t) \equiv \Theta - \Theta_o = - \frac{\pi}{\tau_o t_o} t^2\label{e5}
\end{equation}
Equation (\ref{e5}) gives the average perhelion precession at time '$t$'.
Specializing to the case of Mercury, where $\tau_o = \frac{1}{4}$ year, it
follows from (\ref{e5}) that the average precession per year at time '$t$' is
given by
\begin{equation}
\omega (t) =  \frac{4\pi t^2}{t_0}\label{e6}
\end{equation}
Whence, considering $\omega (t)$ for years $t=1,2, \cdots , 100,$ we can
obtain from (\ref{e6}), the usual total perhelion precession per century as,
$$\omega = \sum^{100}_{n=1} \omega (n) \approx 43'' ,$$
if the age of the universe is taken to be $\approx 2 \times 10^{10}$ years.\\
Conversely, if we use the observed value of the precession in (\ref{e6}),
we can get back the above age of the universe.\\
It can be seen from (\ref{e6}), that the precession depends on the epoch.\\
We next demonstrate that orbiting objects will have an anamolous inward
radial acceleration.\\
Using the well known equation for Keplarian orbits (cf.ref.\cite{r7}),
\begin{equation}
\frac{1}{r} = \frac{GMm^2}{l^2} (1 + e cos \Theta)\label{e7}
\end{equation}
\begin{equation}
\dot r^2 = \frac{GM}{rm} - \frac{l^2}{m^2r^2}\label{e8}
\end{equation}
$l$ being the orbital angular momentum constant and $e$ the eccentricity of
the orbit, we can deduce such an extra inward radial acceleration, on
differentiation of (\ref{e8}) and using (\ref{e2}) and (\ref{e3}),
\begin{equation}
a_r = \frac{GM}{2t_o r \dot r}\label{e9}
\end{equation}
It can be easily shown from (\ref{e7}) that
\begin{equation}
\dot r \approx \frac{eGM}{rv}\label{e10}
\end{equation}
For a nearly circular orbit $rv^2 \approx GM$, whence use of (\ref{e10}) in
(\ref{e9}) gives,
\begin{equation}
a_r \approx v/2 t_o e\label{e11}
\end{equation}
For the earth, (\ref{e11}) gives an anomalous inward radial acceleration
$\sim 10^{-9} cm/sec^2,$ which is known to be the case\cite{r8}.\\
We could also deduce a progressive decrease in the eccentricity of orbits. Indeed,
$e$ in (\ref{e7}) is given by
$$e^2 = 1+\frac{2El^2}{G^2m^3M^2} \equiv 1 + \gamma , \gamma < 0.$$
Use of (\ref{e2}) in the above and differenciation, leads to,
$$\dot e = \frac{\gamma}{et_o} \approx - \frac{1}{et_o} \approx - \frac{10^{-10}}
{e} \mbox{per year},$$
if the orbit is nearly circular. (Variation of eccentricity in the usual
theory have been extensively studied (cf.ref.\cite{r9} for a review).\\
We finally consider the anomalous accelerations given in (\ref{e9}) and (\ref{e11})
in the context of space crafts leaving the solar system.\\
If in (\ref{e9}) we use the fact that $\dot r \leq v$ and approximate
$$v \approx \sqrt{\frac{GM}{r}},$$
we get,
$$a_r \geq \frac{1}{et_o} \sqrt{\frac{GM}{r}}$$
For $r \sim 10^{14}cm$, as is the case of Pioneer $10$, this gives,
$a_r \geq 10^{-11}cm/sec^2$\\
Interestingly Anderson et al.,\cite{r10} claims to have observed an anomalous inward acceleration
of $\sim 10^{-9} cm/sec^2$

\end{document}